\documentclass{aa}
\usepackage{graphicx}
\usepackage{txfonts}
\begin{document}

  \title{Multifrequency VLBA Monitoring of 3C\,273 during the INTEGRAL Campaign in
  2003}

  \subtitle{I. Kinematics of the Parsec Scale Jet from 43 GHz Data}

  \author{T. Savolainen
          \inst{1}
	  \and
          K. Wiik
	  \inst{1,2}
	  \and
	  E. Valtaoja
	  \inst{1,3}
	  \and
	  M. Tornikoski
	  \inst{4}}

  \offprints{T. Savolainen}

  \institute{Tuorla Observatory, 
	     V\"ais\"al\"antie 20, FI-21500, Piikki\"o, Finland\\
	     \email{tukasa@utu.fi}
	     \and
	     Institute of Space and Astronautical Science, Japan Aerospace
	     Explorations Agency, 3-1-1 Yoshindai, Sagamihara, 
	     Kanagawa 229-8510, Japan
	     \and
	     Department of Physics, University of Turku, FI-20014 Turku, 
	     Finland
	     \and 
	     Mets\"ahovi Radio Observatory, Helsinki University
	     of Technology, Mets\"ahovintie 114, FI-02540
	     Kylm\"al\"a, Finland 
}

  \date{Received 4 July 2005; accepted 29 August 2005}

  \abstract{In this first of a series of papers describing
    polarimetric multifrequency Very Long Baseline Array (VLBA)
    monitoring of \object{3C\,273} during a simultaneous campaign with
    the INTEGRAL $\gamma$-ray satellite in 2003, we present 5 Stokes I
    images and source models at 7~mm. We show that a part of the inner
    jet (1--2 milliarcseconds from the core) is resolved in a
    direction transverse to the flow, and we analyse the kinematics of
    the jet within the first 10~mas. Based on the VLBA data and
    simultaneous single-dish flux density monitoring, we determine an
    accurate value for the Doppler factor of the parsec scale jet, and
    using this value with observed proper motions, we calculate the
    Lorentz factors and the viewing angles for the emission components
    in the jet. Our data indicates a significant velocity gradient
    across the jet with the components travelling near the southern
    edge being faster than the components with more northern path. We
    discuss our observations in the light of jet precession model and
    growing plasma instabilities.

   }

  \titlerunning{Multifrequency VLBA Monitoring of 3C\,273 in 2003. I}
  \maketitle

%

\section{Introduction}

Although vigorously investigated for over forty years, the physics of
active galactic nuclei (AGN) have proved to be far from easily
understandable. A set of basic facts about the nature of the central
engine being a supermassive black hole and the gravitational accretion
process as the primary energy source have been established beyond
reasonable doubt, but the details of the picture do not quite fit. The
intricate physics of AGN is made difficult by the complex interplay
between several emission mechanism working simultaneously. For
example, blazars, the class unifying flat-spectrum radio-loud quasars
and BL Lac objects, contains sources that show significant radiation
throughout the whole electromagnetic spectrum from radio to TeV
$\gamma$-rays. The different emission mechanisms in AGNs include e.g.
non-thermal synchrotron radiation from a relativistic jet, thermal
emission from dust, optical-ultraviolet emission from the accretion
disk, Comptonized hard X-rays from the accretion disk corona and
$\gamma$-rays produced by inverse Compton scattering in the
jet. Therefore, observing campaigns monitoring the sources
simultaneously at as many wavelengths as possible are essential in
studying AGNs.

Since the relativistic jet plays a crucial role in the objects
belonging to the blazar class of AGNs, Very Long Baseline
Interferometry (VLBI) observations imaging the parsec scale jet are an
important addition to campaigns concentrating on this class. In
particular, combined multifrequency VLBI monitoring and $\gamma$-ray
measurements with satellite observatories may provide a solution to
the question of the origin of the large amount of energy emitted in
$\gamma$-rays by some blazars. By studying time lags between
$\gamma$-ray flares and ejections of superluminal components into the
VLBI jet, we should be able to establish whether the high energy
emission comes from a region close to the central engine or whether it
instead originates in the radio core or in the relativistic shocks --
parsecs away from the black hole. In addition, multifrequency VLBI
data yields spectra of the radio core and superluminal knots, which
can be used to calculate the anticipated amount of the high energy
emission due to the synchrotron self-Compton (SSC) process if an
accurate value of the jet Doppler factor is known. Predicted SSC-flux
can be directly compared with simultaneous hard X-ray and $\gamma$-ray
observations.

Due to its proximity, \object{3C\,273} (z=0.158; Schmidt \cite{sch63})
-- the brightest quasar on the celestial sphere -- is one of the most
studied AGNs (see Courvoisier \cite{cou98} for a comprehensive
review). Because \object{3C\,273} exhibits all aspects of nuclear
activity including a jet, a blue bump, superluminal motion, strong
$\gamma$-ray emission and variability at all wavelengths, it was
chosen as the target of an INTEGRAL campaign, where radiation above
1~keV was monitored in order to study the different high energy
emission components (Courvoisier et al. \cite{cou03}). INTEGRAL
observed \object{3C\,273} in January, June-July 2003, and January 2004
and the campaign was supported by ground based monitoring at lower
frequencies carried out at several different observatories
(Courvoisier et al. \cite{cou04}). This article is the first in a
series of papers describing a multifrequency polarimetric VLBI
monitoring of \object{3C\,273} using the Very Long Baseline
Array\footnote{The VLBA is a facility of the National Radio Astronomy
  Observatory, operated by Associated Universities, Inc., under
  cooperative agreement with the U.S. National Science Foundation.}
(VLBA) in conjunction with the INTEGRAL campaign during 2003. The
amount of data produced by such a VLBI campaign using six frequencies
is huge. In order to divide the large volume of the gathered
information into manageable chunks, we will treat the different
aspects of the data (i.e. kinematics, spectra and polarisation) in
different papers, and in the end we will attempt to draw a
self-consistent picture taking into account the whole data set.

In the current paper we construct a template for the later analysis of
the component spectra and polarisation, and for the comparison of the
VLBI data with the INTEGRAL measurements. Since kinematics is
important in all subsequent analyses, we start by studying the
component motion in the jet within 10~mas from the radio core. The
emphasis is on the components located close to the core ($\lesssim
2$~mas), since they are the most relevant for the combined analysis of
the VLBI and INTEGRAL data. We use the 43~GHz data for the kinematical
study, since they provide the best angular resolution with a good
signal-to-noise ratio (SNR). The other frequencies are not discussed
in this paper, since the positional errors at all other frequencies
are significantly larger for the components we are interested in
(i.e. those located near the core). Also, a kinematical analysis that
takes into account multiple frequencies in a consistent way is very
complicated using contemporary model fitting methods. However, the
other frequencies are not ignored either; instead, we will check the
compatibility of this paper's results with the multifrequency data
presented in the forthcoming papers treating the spectral
information. In Paper~II (Savolainen et al. in preparation) we will
discuss the method to obtain the spectra of individual knots in the
parsec scale jet, and present the spectra from the February 2003
observation. We will calculate the anticipated amount of
SSC-radiation, and compare this with the quasi-simultaneous INTEGRAL
data. In Papers III and IV, spectral evolution and polarisation
properties of the superluminal components over the monitoring period
will be presented. Paper~V combines all the obtained information, to
interpret it in a self-consistent way.

Throughout the paper, we use a contemporary cosmology with the
following parameters: $H_0$ = 71~km~s$^{-1}$~Mpc$^{-1}$, $\Omega_M$ =
0.27 and $\Omega _\Lambda$ = 0.73. For the spectral index, we use the
positive convention: $S_{\nu} \propto \nu^{+\alpha}$.

%

\section{Observations and data reduction}
We observed \object{3C\,273} with the National Radio Astronomy
Observatory's Very Long Baseline Array at five epochs in 2003
(February 28th, May 11th, July 2nd, September 7th and November 23rd)
for nine hours at each epoch, using six different frequencies (5, 8.4,
15, 22, 43 and 86~GHz). The observations were carried out in the
dual-polarisation mode enabling us also to study the linear
polarisation of the source. However, we do not discuss the
polarisation data here; it will be published in Paper~IV (Savolainen
et al. in preparation). At the first two epochs, \object{3C\,279} was
observed as a polarisation calibrator and as a flux calibrator. At
the three later epochs, we added two more EVPA calibrators to the
program: \object{OJ\,287} and \object{1611+343}.

The data were calibrated at Tuorla Observatory using NRAO's
Astronomical Image Processing System (AIPS; Bridle \& Greisen
\cite{bri94}) and the Caltech Difmap package (Shepherd \cite{she97})
was used for imaging. Standard procedures of calibration, imaging and
self-calibration were employed, and final images were produced with
the Perl library
FITSPlot\footnote{http://personal.denison.edu/$^\sim$homand/}.

Accuracy of {\it a priori} flux scaling was checked by comparing
extrapolated zero baseline flux densities of a compact calibrator
source, \object{3C\,279}, to near-simultaneous observations from VLA
polarisation monitoring program (Taylor \& Myers \cite{tay00}) and
from Mets\"ahovi Radio Observatory's quasar monitoring program
(Ter\"asranta et al. \cite{ter04}). The comparison shows that the {\it
a priori} amplitude calibration of our data at 43~GHz is accurate to
$\sim 5$\%, which is better than the often quoted nominal value of
10\%.  A detailed discussion of the amplitude calibration and setting
of the flux scale will be given in Paper~II.

The angular resolution of the VLBI data is a function of observing
frequency, {\it uv}-coverage and signal-to-noise ratio. Only 7
antennas out of 10 were able to observe at 86~GHz during our campaign
(Brewster and St. Croix did not have 3~mm receivers, and the baselines
to Hancock did not produce fringes at any observation epoch), and
hence, the longest baselines (Mauna Kea -- St. Croix, Mauna Kea --
Hancock) were lost. This, combined with a significantly lower SNR,
results in poorer angular resolution at 86~GHz than at 43~GHz, and
thus, unfortunately, 3~mm data does not provide any added value for
the kinematical study as compared to 43~GHz. (However, for the
spectral analysis 3~mm data is invaluable, as will be shown in
Paper~II.) At lower frequencies, the {\it uv}-coverage is identical to
the one at 43~GHz, and, although the SNR is better, the eventually
achieved angular resolution in model fitting is poorer. Currently,
there is no model fitting procedure capable of fitting multiple
frequencies simultaneously, which makes utilising multifrequency data
in the kinematical analysis quite difficult due to different
positional uncertainties at different frequencies, and due to a
frequency dependent core position. For these reasons, the kinematical
analysis, which provides the needed template for further study of
spectra and polarisation of the radio core and the components close to
it, is done at the single frequency maximising the angular resolution
i.e. 43~GHz. Obviously, the results are preliminary in that the later
analysis of the spectra and polarisation will provide a check on the
kinematics presented in this paper.

\subsection{Modelling and estimation of model errors}

\begin{figure}
\resizebox{0.96\hsize}{!}{\includegraphics{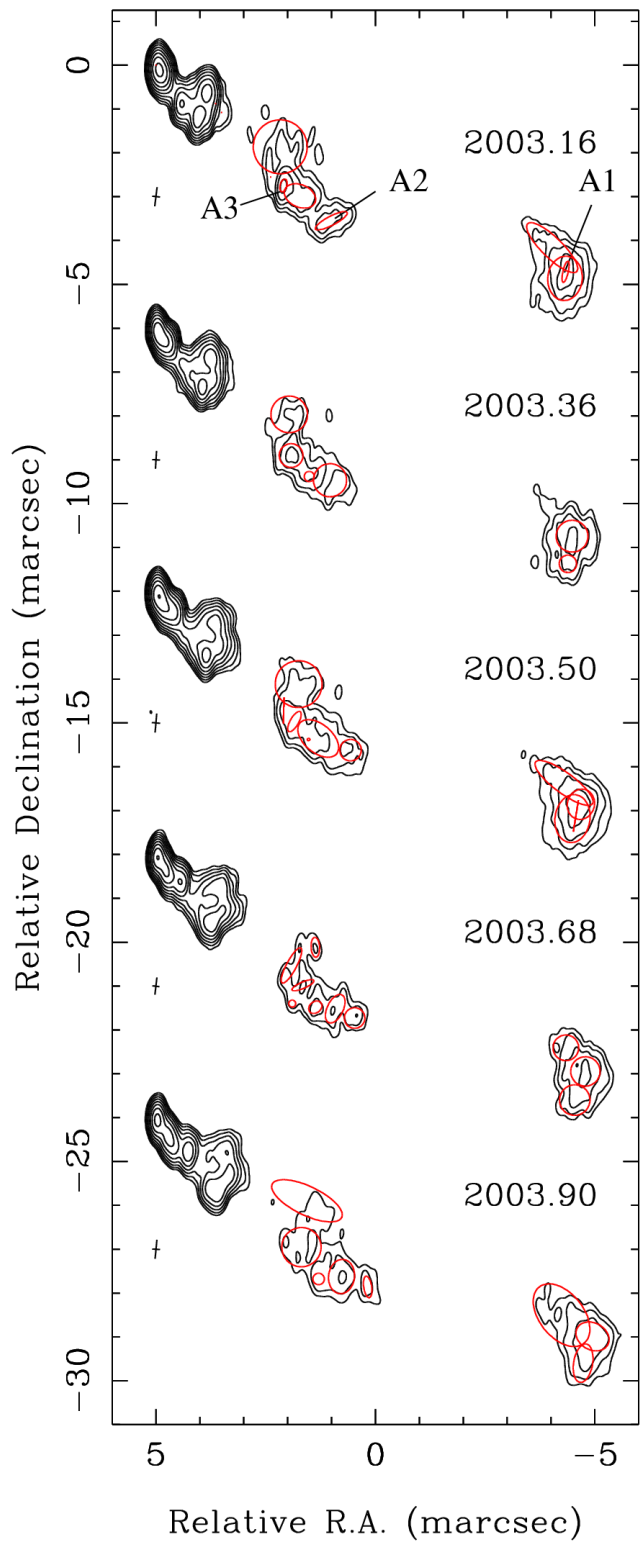}}
\caption{Naturally weighted final CLEAN images of \object{3C\,273} at
43~GHz. Contour levels are represented by geometric series
$c_0(1,...,2^n)$, where $c_0$ is the lowest contour level
corresponding to 5 times the off-source rms noise, which is given in
Table~\ref{clean_param} together with the restoring beam size and peak
intensity for each image. Overlaid on the contour images are model-fit
components in the outer part of the jet, and three prominent
components are marked with A1--A3. The model-fit components located
within the inner 2.5~mas from the core are shown in
Fig.~\ref{model_images}. See the on-line edition for a colour
version of the figure.}
\label{clean_images}
\end{figure}

\begin{table*}
\caption{Parameters of the images in Fig.~1}
\label{clean_param}
\centering
\begin{tabular}{cccccc} \hline \hline
Epoch & $\Theta_{b,\textrm{\scriptsize{maj}}}$ & $\Theta_{b,\textrm{\scriptsize{min}}}$ & 
P.A. & $\sigma(I)$ & $I_{\textrm{\scriptsize{peak}}}$ \\ 
 {[yr]} & [mas] & [mas] & [$\degr$] & [mJy beam$^{-1}$] & [mJy beam$^{-1}$] \\ \hline
2003.16 & 0.44 & 0.18 & -7.1 & 2.4 & 3961 \\
2003.36 & 0.40 & 0.18 & -3.5 & 2.2 & 2750 \\ 
2003.50 & 0.43 & 0.19 & -4.6 & 2.0 & 2600 \\
2003.68 & 0.39 & 0.17 & -6.3 & 1.7 & 2241 \\
2003.90 & 0.44 & 0.18 & -5.5 & 2.4 & 1997 \\ \hline
\end{tabular}
\end{table*}

\begin{figure*}
\sidecaption
 \includegraphics[width=12cm]{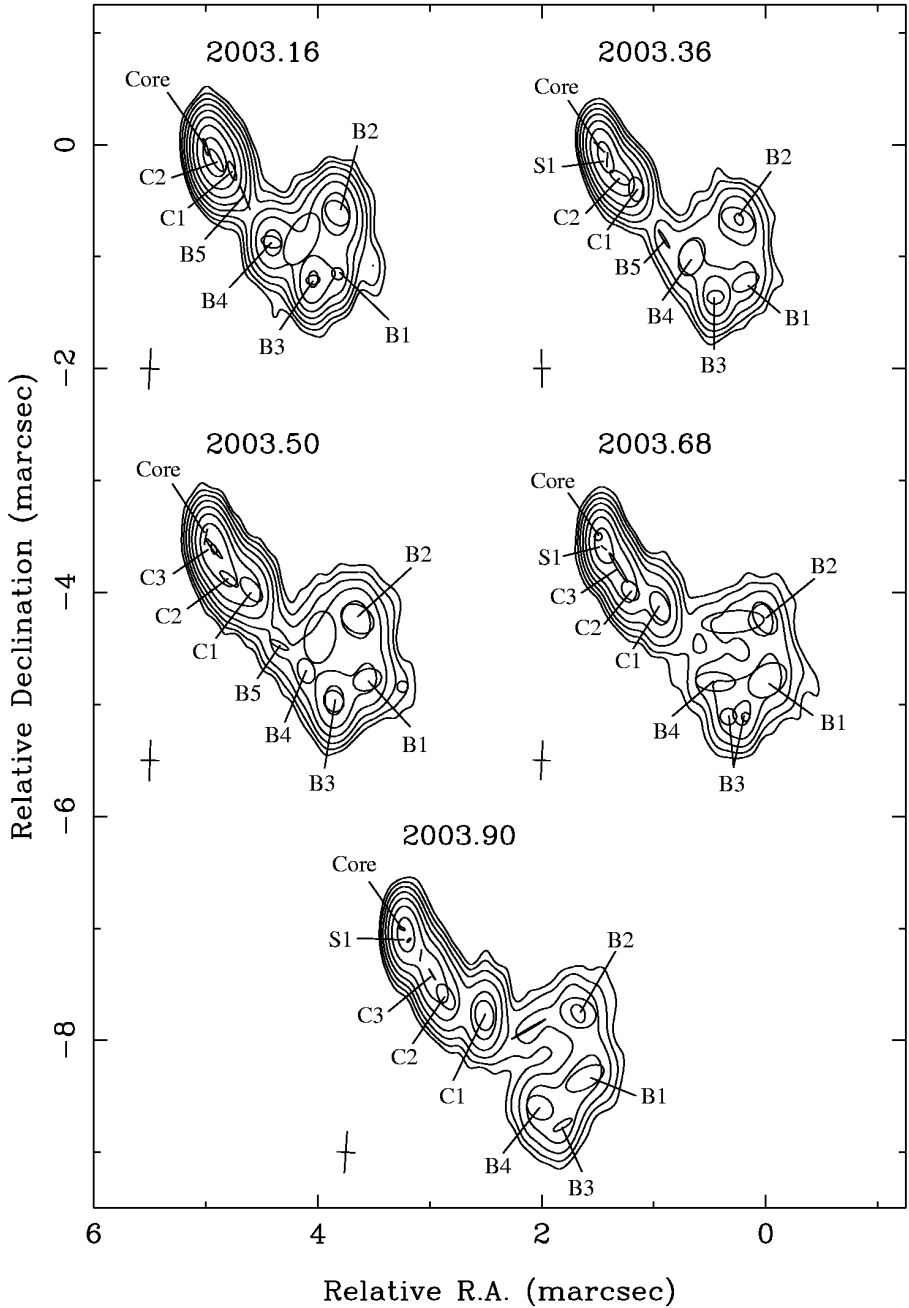}
 \caption{~Gaussian model components within 2.5~mas from the core and
 their convolution with a Gaussian beam corresponding to uniformly
 weighted ($u$,$v$) data with residuals added. Table~\ref{model_param}
 lists restoring beam sizes, peak intensities and off-source rms
 noise levels. Contour levels are represented by geometric series
 $c_0(1,...,2^n)$, where $c_0$ is the lowest contour level
 corresponding to 5 times the off-source rms noise.}
 \label{model_images}
\end{figure*}

The final CLEAN images of \object{3C\,273} (Fig.~\ref{clean_images})
are not easy to parameterise for quantitative analysis of e.g. proper
motions and flux density evolution of different regions in the jet. To
make a quantitative analysis feasible, we fitted models consisting of
a number of elliptical and circular Gaussian components to fully
self-calibrated visibility data (i.e. in the ($u$,$v$)-plane) using
the ``modelfit'' task in Difmap. We sought to obtain the simplest
model that gave a good fit to the visibilities and produced, when
convolved with the restoring beam, a brightness distribution similar
to that of the CLEAN image. \object{3C\,273} is a complex source with
a number of details, and due to its structure that could not easily be
represented by well-separated, discrete components, model-fitting
proved to be rather difficult. Due to this complexity, we added more
constraints to the fitting process by demanding consistency between
epochs, i.e. most components should be traceable over time. The final
models show a good fit to the visibility data, and they nicely
reproduce the structure in the CLEAN images (compare
Figs.~\ref{clean_images} and \ref{model_images}, the first showing
CLEAN images and the latter presenting model components convolved with
a Gaussian beam). However, these models (like any result from the
model-fitting of the VLBI data) are not unique solutions, but rather
show one plausible representation of the data.

Reliable estimation of uncertainties for parameters of the model
fitted to VLBI data is a notoriously difficult task. This is due to
loss of knowledge about the number of degrees of freedom in the data,
because the data are no longer independent after self-calibration is
applied. This makes the traditional $\chi^2$-methods useless, unless a
well-justified estimate of degrees of freedom can be made. Often there
is also strong interdependency between the components -- for instance,
adjacent components' sizes and fluxes can be highly dependent on each
other. This can be solved by using a method first described by
Tzioumis et al. (\cite{tzi89}), where the value of the model parameter
under scrutiny is adjusted by a small amount and fixed. The
model-fitting algorithm is then run and results are inspected. This
cycle is repeated until a clear discrepancy between the model and the
visibilities is achieved. {\it Difwrap} (Lovell \cite{lov00}) is a
program to automatically carry out this iteration, and we have used it
to assess uncertainties for parameters of the model
components. However, there is a serious problem also with this
approach. Since the number of degrees of freedom remains unknown, we
cannot use any statistically justified limit to determine what is a
significant deviation from the data; we must rely on highly subjective
visual inspection of the visibilities and of the residual map. We
found that a reasonable criterion for a significant discrepancy
between the data and model is the first appearance of such an emission
structure, which we would have cleaned, had it appeared during the
imaging process. For faint components, the parameters can change
substantially with insignificant increase in the difference between
the model and the visibilities, and hence, with low flux components we
cannot rely on the results given by Difwrap, but we need to use other
means to estimate the uncertainties. The above-described procedure for
determination of uncertainties in model parameters does not address
errors arising from significant changes in the model (e.g. changes in
the number of components) or from imaging artefacts created by badly
driven imaging/self-calibration iteration.

Homan et al. (\cite{hom01}) estimated the positional errors of the
components by using the variance about the best-fit proper motion
model. We applied their method to our data in order to estimate the
positional uncertainties of weak components and to check the
plausibility of the errors given by Difwrap for stronger components.
We have listed the average positional error $\langle\Delta
r_{DW}\rangle$ from the Difwrap analysis and the rms variation about
the best-fit proper motion model $\sigma (r)$ for each component that
could be reliably followed between the epochs in
Table~\ref{errors}. The errors are comparable -- for components
B1--B4\footnote{The naming convention of the components in this paper
is our own and does not follow any previous practice.} they are almost
identical and for components C1--C3 the uncertainties obtained by
Difwrap are 0.01--0.02~mas larger than the rms fluctuations about the
proper motion model. Diffuse components A1, A2 and A3 have Difwrap
positional errors approximately twice as large as the variance about
the best-fit model. We conclude that Difwrap yields reliable estimates
for the positional uncertainties of the compact and rather bright
components. As was mentioned earlier, it is difficult to obtain an
error estimate for weak components with this method, and for diffuse
components, the errors seem to be exaggerated. The average errors
given in Table~\ref{errors} agree well with the often quoted ``$1/5$
of the beam size'' estimate for positional accuracy of the model
components (e.g. Homan et al. \cite{hom01}, Jorstad et
al. \cite{jor05}).

\begin{table}
\caption{Average positional uncertainties for the model components
from the Difwrap ($\langle\Delta r_{DW}\rangle$) and from the variance analysis ($\sigma (r)$)}
\label{errors}
\centering
\begin{tabular}{ccc} \hline \hline
Component & $\langle\Delta r_{DW}\rangle$ & $\sigma (r)$ \\ 
 & [mas] & [mas] \\ \hline
C1 & 0.03 & 0.01 \\
C2 & 0.04 & 0.02 \\
C3 & 0.04 & 0.02 \\
B1 & 0.04 & 0.04 \\
B2 & 0.03 & 0.03 \\
B3 & 0.05 & 0.05 \\
B4 & 0.05 & 0.03 \\
B5 & --   & 0.06 \\
A1 & 0.10 & 0.05 \\
A2 & 0.11 & 0.06 \\
A3 & 0.06 & 0.02 \\ \hline
\end{tabular}
\end{table}

Uncertainty in the flux density of a Gaussian component strongly
depends on possible ``flux leakage'' between adjacent components,
which is, in principle, taken into account in the Difwrap
analysis. This flux leakage results in large and often asymmetric
uncertainties, as can be seen in Figs.~\ref{Cflux} and \ref{Bflux},
where flux density evolution of the individual components is
presented. The error bars in the figures correspond to the total
uncertainty in the component flux density, where we have quadratically
added the model error given by Difwrap and a 5\% amplitude calibration
error. The errors are on average 10\% for moderately bright and rather
isolated components with no significant flux leakage from other
components, and $\sim 15$\% for components with the leakage
problem. For some bright and well-isolated components, flux density
errors less than 10\% were found, but in general uncertainties
reported here are larger than those obtained by the $\chi^2$-method,
which does not take into account the interdependency between nearby
components' flux densities (e.g. Jorstad et
al. \cite{jor01}). Naturally, the caveat in the Difwrap method is the
subjective visual inspection needed to place the error boundaries, but
we note that Homan et al. (\cite{hom02}), who estimated flux density
errors by examining the correlation of flux variations between two
observing frequencies and taking an uncorrelated part as a measure of
uncertainty, found flux density errors ranging from 5\% to 20\% at
22~GHz with 5\%--10\% being typical values for well-defined
components.  Our error estimates for component flux densities seem to
agree with these values.

Uncertainties in the component sizes were also determined with
Difwrap. Plausible values for errors were found only for bright
components, and again, asymmetric uncertainties emerged commonly. The
uncertainties in the size of the major axis of the component ranged
from 0.01~mas to 0.15~mas corresponding to approximately 5\%--30\%
relative uncertainty. Errors in the component axial ratios were
difficult to determine with Difwrap, and results were obtained only
for a small number of components. The axial ratio uncertainties range
from 0.05 to 0.20, corresponding to a relative uncertainty of about
20\%--60\%. The large errors in the component sizes are again due to
``leakage'' between adjacent components. For well-isolated components,
uncertainties in size would be smaller.

%

\section{The structure of the parsec scale jet}

\begin{table*}
\caption{Parameters of the images in Fig. 2}
\label{model_param}
\centering
\begin{tabular}{cccccc} \hline \hline
Epoch & $\Theta_{b,\textrm{\scriptsize{maj}}}$ & $\Theta_{b,\textrm{\scriptsize{min}}}$ & 
P.A. & $\sigma(I)$ & $I_{\textrm{\scriptsize{peak}}}$ \\ 
 {[yr]} & [mas] & [mas] & [$\degr$] & [mJy beam$^{-1}$] & [mJy beam$^{-1}$] \\ \hline
2003.16 & 0.37 & 0.17 & -3.1 & 4.0 & 3833 \\
2003.36 & 0.34 & 0.16 & 1.0 & 6.4 & 2605 \\ 
2003.50 & 0.37 & 0.17 & -1.8 & 3.7 & 2432 \\
2003.68 & 0.35 & 0.16 & -3.1 & 4.7 & 2242 \\
2003.90 & 0.38 & 0.16 & -3.7 & 4.1 & 1918 \\ \hline
\end{tabular}
\end{table*}

In our 43~GHz, high dynamic range images (Fig.~\ref{clean_images}; see
also Table~\ref{clean_param}) we can track the jet up to a distance of
$\sim 11$~mas from the core. The ``wiggling'' structure reported in
several other studies (e.g. Mantovani et al. \cite{man99}, Krichbaum
et al. \cite{kri90}, B{\aa}{\aa}th et al. \cite{baa91}, Zensus et
al. \cite{zen88}) is evident also in our images -- the bright features
are not collinear.

\begin{figure}
\resizebox{\hsize}{!}{\includegraphics{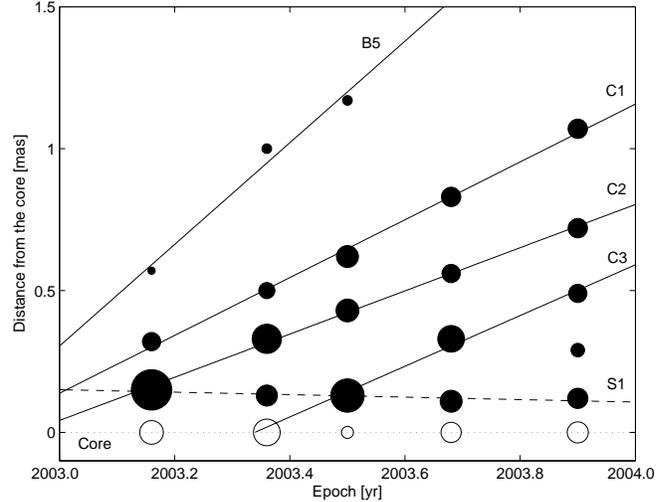}}
\caption{Separation of the components from the core vs. epoch of
observation for the components located within 1.5~mas from the
core. The size of the circle is proportional to square root of the
flux density. The solid lines represent the best linear approximation
of the data for moving components, while the dashed line is the
best-fit to the positions of stationary component S1. We used linear
fits here for all components, although C1 is better fitted with a
second-order polynomial as shown in Fig.~\ref{Cmotion}. The open
circles correspond to the core.}
\label{Cflux_motion}
\end{figure}

At all epochs, we consider the compact, easternmost component as the
stationary radio core, and all proper motions are measured relative to
it.  As can be seen from Fig.~\ref{model_images}, the innermost region
``C'' with the core and several departing components (C1--C3) extends
about 1~mas to south-west. Besides the core, there is also another
stationary component, S1, located $\sim0.15$~mas from the
core. Identification of the components close to the core over the
epochs is problematic due to the number of knots close to each other,
and also because new ejections are going on during our observing
period. We have plotted the components within 1.5~mas from the core in
Fig.~\ref{Cflux_motion} with the size of the symbol indicating the
relative flux of each Gaussian component. The figure shows four moving
components (C1--C3 and B5), and two stationary components (the core
and S1), with consistent flux evolution. The figure also confirms that
component C3 is ejected from the core during our monitoring period.
Components C1--C3 may not necessarily represent independent and
distinct features, but they also can be interpreted as a brightening
of the inner jet -- e.g. due to continuous injection of energetic
particles in the base of the jet. VLBI data with better angular
resolution could settle this, but, unfortunately, our 86~GHz
observations suffer from the loss of the longest VLBA baselines, and
thus, they cannot provide the necessary resolution. However, this
should not affect the following kinematical analysis.

In the complex emission region at about 1--2.5~mas to southwest from
the core, the jet is resolved in the direction transverse to the
flow. We have identified five components (B1--B5) that are traceable
over the epochs. The components can be followed from epoch to epoch in
Fig.~\ref{model_images}. B3 is represented by two Gaussian components
at the fourth epoch, and B5 is not seen after the third epoch, but
otherwise we consider that our identification is robust. B4 seems to
catch up with the slower knot B3 and collides with it (see
Figs.~\ref{Bxy} and \ref{Bmotion}).

The jet broadens significantly after $\sim1.5$~mas and, as shown later
in Sect.~4, the trajectory of the component B1 does not extrapolate to
the core (Fig.~\ref{model_images}). We discuss the possible
explanations for this anomalous region ``B'' in Sect.~6. Farther down
the jet, there are two more emission regions visible in the 43~GHz
images, both of them showing a curved structure
(Fig.\ref{clean_images}). The first one is located 3--6~mas southwest
of the core and we have identified two components A2 and A3 in
it. This feature likely corresponds to knots B1, bs, b1, B2 and b2 of
Jorstad et al. (\cite{jor05}), with our A2 being their B1/bs, and our
A3 being their B2. However, this cross-identification is not certain,
since the region is complex and, as described in Jorstad et
al. (\cite{jor05}), it contains components with differing speeds and
directions as well as secondary components connected to the main
disturbances. The second emission region in the outer part of the jet
is located at 9--11~mas from the core, and we call this component with
a curved shape A1 (Fig.~\ref{clean_images}).

%

\section{Kinematics of the jet}

\begin{figure}
\resizebox{\hsize}{!}{\includegraphics{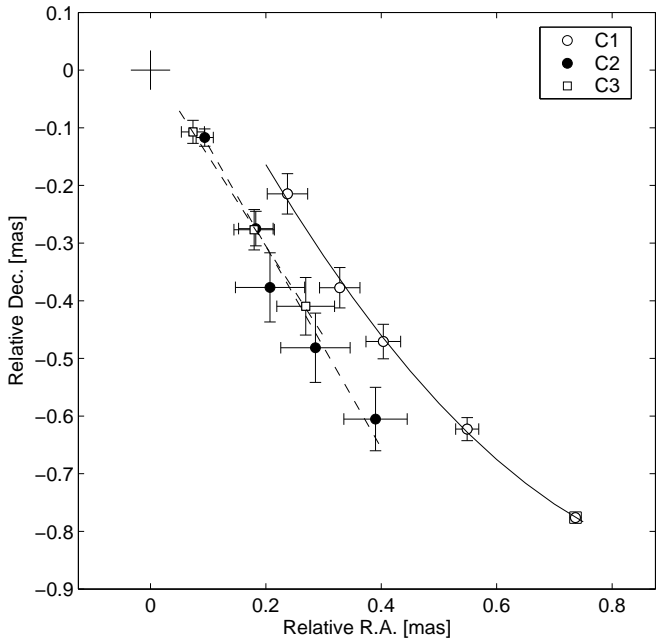}}
\caption{Trajectories of components C1--C3. The symbols show the
measured positions of the components. The solid line is the best-fit
curve for component C1 and the dashed lines show linear trajectories
fitted for components C2 and C3. The plus sign marks the core.}
\label{Cxy}
\end{figure}

\begin{figure}
\resizebox{\hsize}{!}{\includegraphics{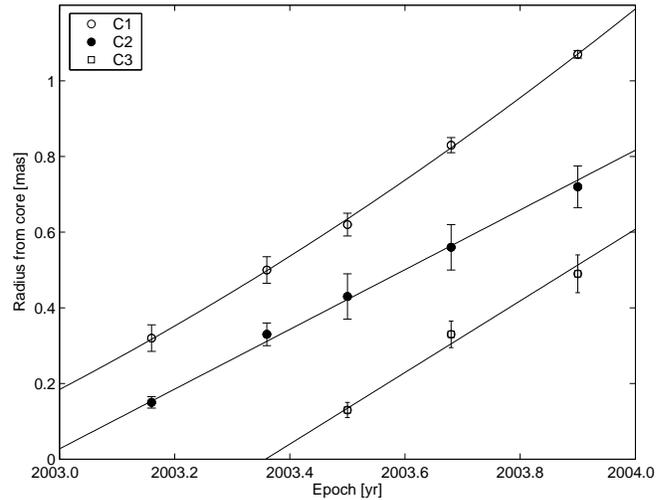}}
\caption{Separation of components C1--C3 from the core vs. epoch of
observation. The solid lines(curve) represent the best-fit straight
lines(second-order polynomial) describing the component motion (see
Table~\ref{proper_motions}).}
\label{Cmotion}
\end{figure}

The data allow us to follow the evolution of the components identified
in the previous section, and to measure their proper motion, which is
the basis for analysing physical properties of the jet. We follow the
practice of Jorstad et al. (\cite{jor05}), and define the average
proper motion as a vector $(\langle\mu\rangle,\langle\Phi\rangle)$,
where
$\langle\mu\rangle=\sqrt{\langle\mu_{\mathrm{RA}}\rangle^2+\langle\mu_{\mathrm{Dec}}\rangle^2}$
is the mean angular speed and
\mbox{$\langle\Phi\rangle=\tan^{-1}\big(\frac{\langle\mu_{\mathrm{RA}}\rangle}{\langle\mu_{\mathrm{Dec}}\rangle}\big)$}
is the direction of motion. In all but one case, the average
coordinate motions $\langle\mu_{\mathrm{RA}}\rangle$ and
$\langle\mu_{\mathrm{Dec}}\rangle$ are obtained by a linear
least-squares fit to the relative (RA,Dec) positions of a component
over the observation epochs. For component C1, a second-order
polynomial was fitted instead of a first-order one, because it reduced
the $\chi^2$ for the RA-coordinate fit from 4.7 (with 3 degrees of
freedom) to 0.2 (with 2 degrees of freedom). Also, the (RA,Dec)-plot
in Fig.\ref{Cxy} indicates a non-ballistic trajectory for component
C1. The amount of acceleration measured for C1 is
$0.21\pm0.18$~mas~yr$^{-2}$ -- a marginal detection
(Fig.~\ref{Cmotion}).

\begin{table*}
\caption{Proper motion results}
\label{proper_motions}
\centering
\begin{tabular}{cccccc} \hline \hline
Component & $S_{\mathrm{max}}$ [Jy] &$\langle\Phi\rangle$ [$^\circ$] & $\langle\mu\rangle$ [mas yr$^{-1}$] & $\langle\beta_{\mathrm{app}}\rangle$ [h$^{-1}$ c] & $T_0$ [yr] \\ \hline
S1  & 1.28 & 30.6 & $0.05\pm0.05$ & $0.4\pm0.4$ & -- \\  
C1  & 1.34 & $-139.4\pm0.3$ & $1.01\pm0.05$ & $7.3\pm0.4$ & $2002.76\pm0.04$ \\
C2  & 4.38 & $-150.5\pm0.7$ & $0.79\pm0.07$ & $5.7\pm0.5$ & $2002.97\pm0.05$ \\ 
C3  & 3.00 & $-147.5\pm1.0$ & $0.95\pm0.13$ & $6.9\pm0.9$ & $2003.36\pm0.03$ \\
B1  & 0.80 & $-115.1\pm0.7$ & $0.71\pm0.06$ & $5.1\pm0.4$ & -- \\
B2  & 1.89 & $-113.3\pm0.6$ & $0.63\pm0.05$ & $4.6\pm0.4$ & $2001.09\pm0.19$ \\
B3  & 1.69 & $-141.2\pm0.5$ & $0.99\pm0.06$ & $7.2\pm0.4$ & $2001.63\pm0.11$ \\
B4  & 1.13 & $-141.3\pm0.4$ & $1.34\pm0.07$ & $9.7\pm0.5$ & $2002.40\pm0.06$ \\
B5  & 0.25 & $-149.6\pm1.1$ & $1.80\pm0.25$ & $13.0\pm1.8$ & $2002.83\pm0.10$ \\ 
A1  & 0.87 & $-116.8\pm1.0$ & $0.95\pm0.13$ & $6.9\pm0.9$ & $1992.2\pm1.5$ \\ 
A2  & 0.32 & $-108.0\pm1.2$ & $1.17\pm0.18$ & $8.5\pm1.3$ & -- \\
A3  & 0.48 & $-126.2\pm2.3$ & $1.07\pm0.32$ & $7.8\pm2.3$ & $1999.4\pm1.2$ \\ \hline
\end{tabular}
\end{table*}

\begin{figure}
\resizebox{\hsize}{!}{\includegraphics{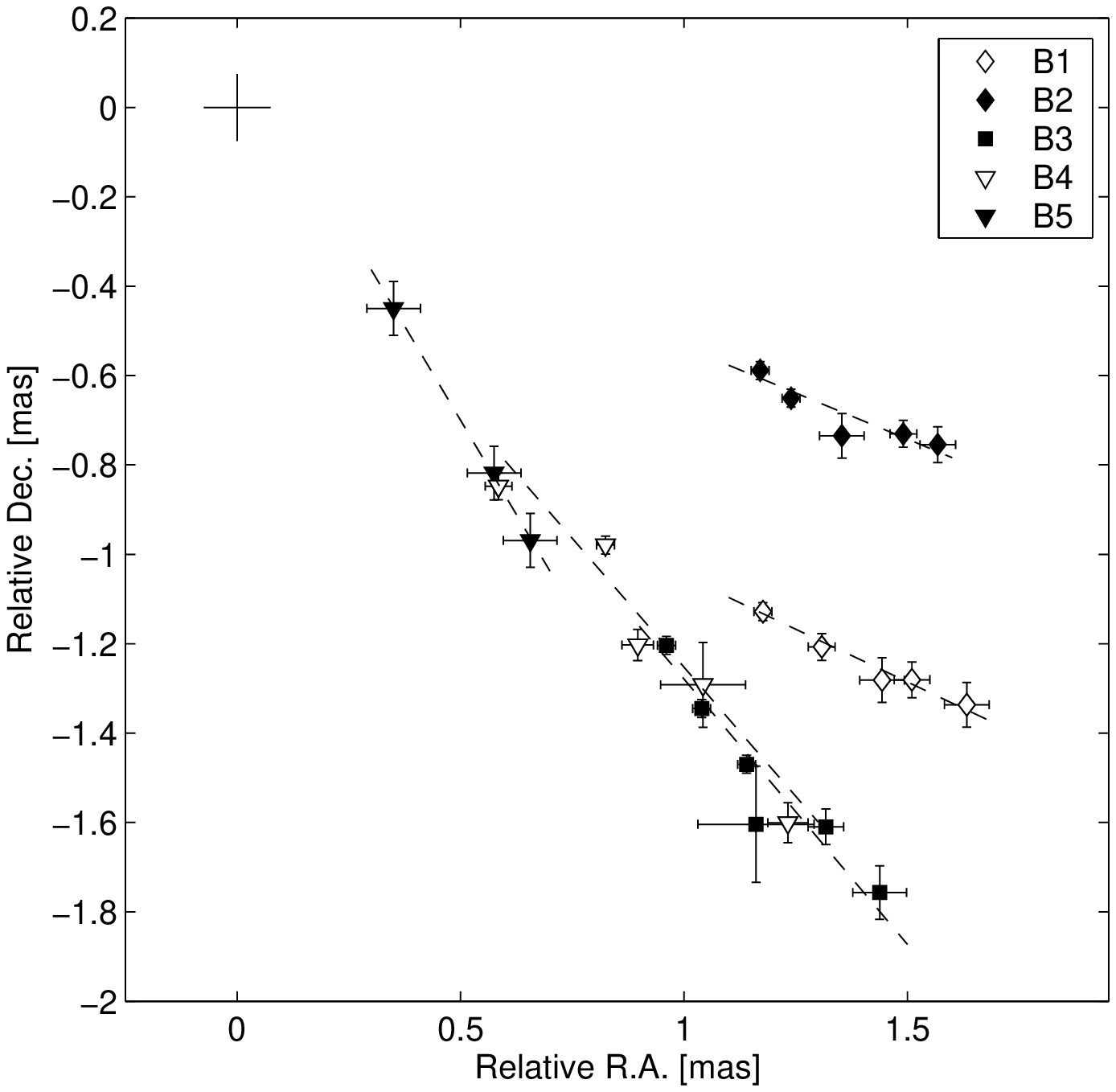}}
\caption{Trajectories of components B1--B5. The symbols show measured
positions of components, and the best-fit linear trajectories are
represented by the dashed lines. The plus sign marks the core. Note
that the trajectory for component B1 clearly does not extrapolate back
to the core.}
\label{Bxy}
\end{figure}

In Figs.~\ref{Cxy}, \ref{Bxy}, and \ref{Axy}, we have plotted the
trajectories in the (RA,Dec)-plane for components C1--C3, B1--B5 and
A1--A3, respectively. As can be seen from these figures, there are
three components showing non-radial motion, C1, B1 and A2. As was
discussed earlier, C1 seems to have a curved trajectory, while B1 and
A2 have straight trajectories, which, however, do not extrapolate back
to the core. B1 has an especially interesting path, since its motion
seems to be parallel to that of B2, but clearly deviating from those
of B3 and B4, which have a direction of $\sim 25 \degr$ more southern
than B1. Also, B1 and B2 have similar speeds, while B3 and B4 are both
faster. B1 might have formed from B2 or B3, possibly through an
interaction between relativistic shocks connected with these
components and the underlying flow, or it might have changed its
trajectory.

\begin{figure}
\resizebox{\hsize}{!}{\includegraphics{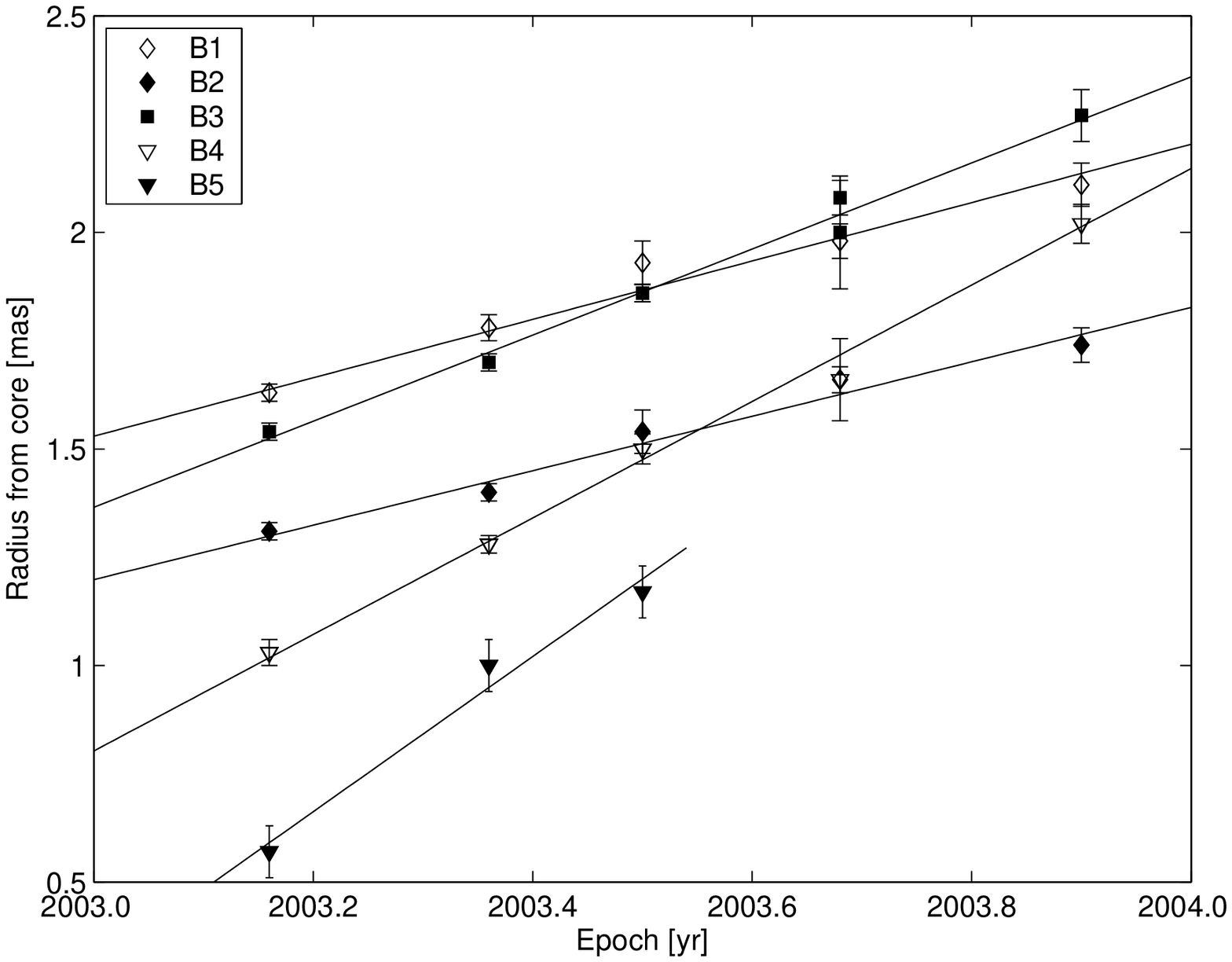}}
\caption{Separation of components B1--B5 from the core vs. epoch of
observation. The solid lines represent the best-fit to the component
motion (see Table~\ref{proper_motions}).}
\label{Bmotion}
\end{figure}

In Figs.~\ref{Cmotion}, \ref{Bmotion}, and \ref{Amotion}, we have
plotted the separation from the core as a function of time for the
components. Table~\ref{proper_motions} lists the maximum measured flux
density, $S_{\mathrm{max}}$, the average proper motion,
$(\langle\mu\rangle, \langle\Phi\rangle)$, the apparent superluminal
speed, $\beta_{\mathrm{app}}$, and the extrapolated epoch of
zero-separation, $T_0$, for the components. Ejection time is not given
for the component S1, which is stationary, nor the components B1 and
A2, whose trajectories do not extrapolate back to the core. The
uncertainties in the average proper motion are derived from the
uncertainties of the fitted polynomial coefficients.

\begin{figure}
\resizebox{\hsize}{!}{\includegraphics{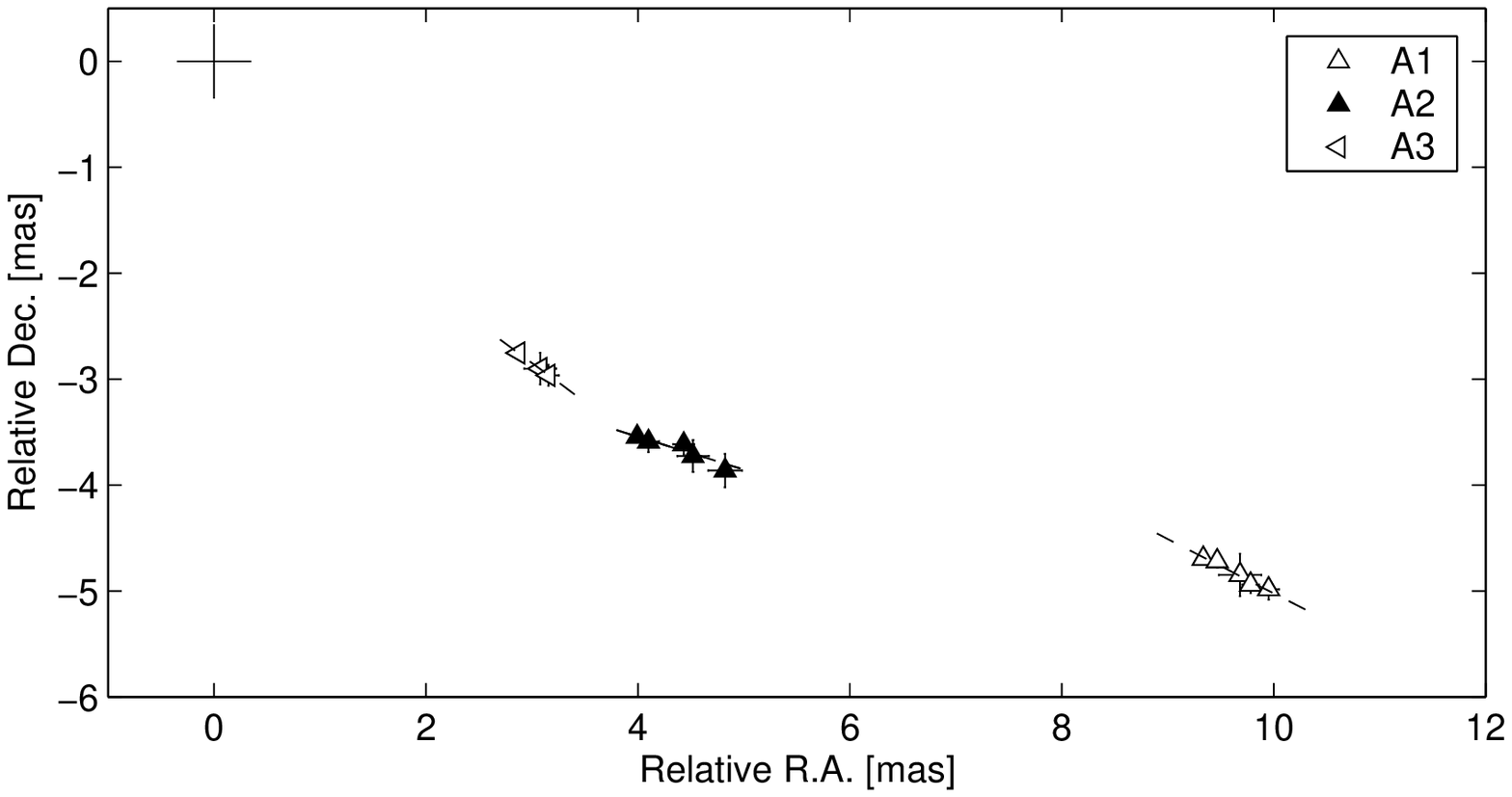}}
\caption{Trajectories of components A1--A3. The symbols show measured
positions of components, and the best-fit linear trajectories are
represented by the dashed lines. The plus sign marks the core.}
\label{Axy}
\end{figure}

\begin{figure}
\resizebox{\hsize}{!}{\includegraphics{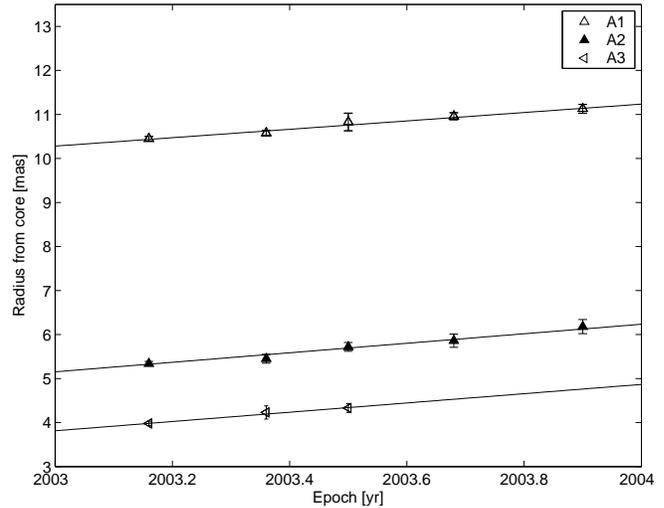}}
\caption{Separation of components A1--A3 from the core vs. epoch of
observation. The solid lines represent the best-fit to the component
motion (see Table~\ref{proper_motions}).}
\label{Amotion}
\end{figure}

The superluminal velocities for the moving components range from
4.6~$h^{-1}$~c to 13.0~$h^{-1}$~c (with $H_0=100h$). This range of
velocities is similar to those previously reported by e.g. Zensus et
al. (\cite{zen90}), Krichbaum et al. (\cite{kri00}), and Jorstad et
al. (\cite{jor05}). The largest observed apparent speed,
13.0~$h^{-1}$~c for B5, is based only on three epochs of observations,
and the component is weak having flux density below 0.25~Jy. This
renders the kinematical result somewhat uncertain, and the high proper
motion of B5 needs to be confirmed by checking against the lower
frequency and polarisation data in the follow-up papers. Components B1
and B2 with $\Phi \approx -115\degr$ have velocities that are
significantly slower than those of the more southern components,
B3--B5, with $\Phi \approx -145\degr$. A similar velocity gradient
transverse to the jet, with components in the northern side having
lower speeds, has been reported by Jorstad et al. (\cite{jor05}).

%

\section{Physical parameters of the jet}

Doppler boosting has an effect on most of the observable properties of
a relativistic jet. Hence, in order to compare observations with
predictions from theory, the amount of Doppler boosting needs to be
measured first. Unfortunately, the Doppler factor of the jet,
\begin{equation}
\delta=[\Gamma(1-\beta \cos \theta)]^{-1}, 
\end{equation}
where $\Gamma=1/\sqrt{1-\beta^2}$ is the bulk Lorentz factor and
$\theta$ is the angle between the jet flow direction and our line of
sight, is difficult to measure accurately (see L\"ahteenm\"aki \&
Valtaoja \cite{lah99b} for a discussion of problems in determining
$\delta$). In this section, we narrow down the possible range of
values for $\delta$, $\Gamma$ and $\theta$ of \object{3C\,273} in
early 2003, so that meaningful comparison with the hard X-ray flux
measured by the INTEGRAL satellite can be made (Paper~II).

The apparent component velocities $\beta_{\mathrm{app}}$ given in
Table~\ref{proper_motions} already put limits on $\Gamma$ and
$\theta$. Namely, for all $\beta_{\mathrm{app}} > 1$ there is a lower
limit for the Lorentz factor $\Gamma_{\mathrm{min}} =
\sqrt{1+\beta_{\mathrm{app}}^2}$ and an upper limit for the viewing
angle, $\sin \theta_{\mathrm{max}} = 2 \beta_{\mathrm{app}} /
(1+\beta_{\mathrm{app}}^2)$. $\Gamma_{\mathrm{min}}$ and
$\theta_{\mathrm{max}}$ for each component are listed in
Table~\ref{gamma_min} (assuming $h=0.71$). These values are calculated
using lower limits of $\beta_{\mathrm{app}}$. As can be seen from
Table~\ref{gamma_min}, the minimum value for $\Gamma$ in the region
close to the VLBI core in 2003 (components C1--C3) is 7.4 and
$\theta_{\mathrm{max}}=15.5^\circ$.

\begin{table}
\caption{Limits for $\Gamma$ and $\theta$}
\label{gamma_min}
\centering
\begin{tabular}{ccc} \hline \hline
Component & $\Gamma_{\mathrm{min}}$ & $\theta_{\mathrm{max}}$ [$\degr$] \\ \hline
C1 & 9.8 & 11.7 \\
C2 & 7.4 & 15.5 \\
C3 & 8.5 & 13.5 \\
B1 & 6.7 & 17.2 \\
B2 & 6.0 & 19.2 \\
B3 & 9.6 & 11.9 \\
B4 & 13.0 & 8.8 \\
B5 & 15.8 & 7.3 \\
A1 & 8.5 & 13.5 \\
A2 & 10.2 & 11.3 \\
A3 & 7.8 & 14.7 \\  \hline
\end{tabular}
\end{table}

\subsection{Variability time scales and Doppler factors}

\begin{figure*}
\sidecaption
 \includegraphics[width=12cm]{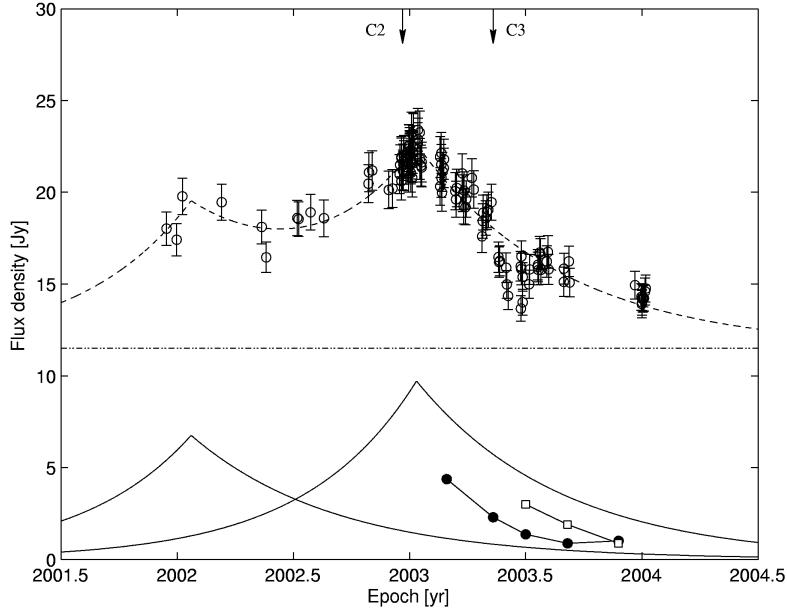}
 \caption{\,Mets\"ahovi total flux density curve of \object{3C\,273}
 at 37~GHz with superposed exponential flare model fits. Open circles
 with error bars correspond to the total flux density measurements,
 the solid curves represent the two fitted flares, the dot-dashed line
 represents a constant baseline flux of 11.5~Jy, and the dashed line
 is the sum of the flares and the baseline. The flux density curves of
 two VLBI components showing flare-like behaviour in 2003, C2 (filled
 circles) and C3 (open squares) are also plotted in the figure. The
 extrapolated ejection epochs of these two components are marked with
 arrows.}
 \label{mhovi_flux}
\end{figure*} 

\begin{figure}
\resizebox{\hsize}{!}{\includegraphics{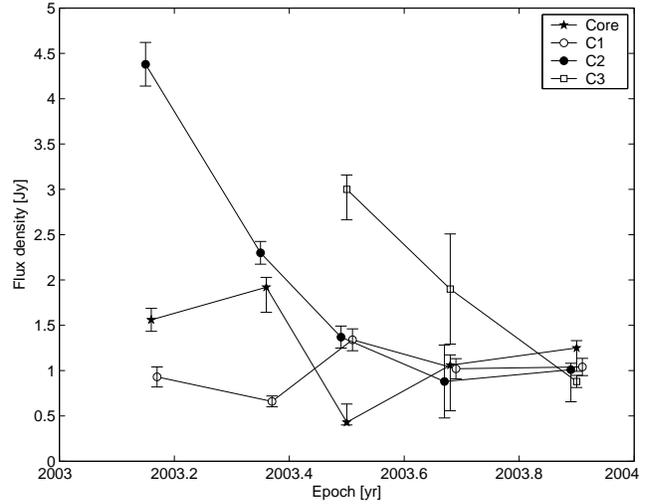}}
\caption{Flux density evolution for the core and components
C1--C3. The asymmetric error bars are mainly due to interdependencies
between flux densities of close-by components.}
\label{Cflux}
\end{figure}

37~GHz total flux density data from the Mets\"ahovi Radio Observatory
monitoring program for 2002--2004 is shown in Fig.~\ref{mhovi_flux}. A
small flare occured at the beginning of 2003. It has about four times
smaller amplitude than the large outburst in 1991 (Robson et
al. \cite{rob93}), making it only a minor but still useful event to
estimate the variability brightness temperature and the Doppler factor
of the jet, as we will show. As is clear from the ejection times
listed in Table~\ref{proper_motions}, components C1--C3 can be
connected to this event. The flux density evolution of C2 (see
Figs.~\ref{mhovi_flux} and \ref{Cflux}) closely matches the flare in
the Mets\"ahovi flux density curve. C1 has an approximately stable
flux density of $\sim 1$~Jy, which makes it a less likely counterpart
to the flare, and C3, on the other hand, was ejected at the time when
the flare was already fading. The flux density of the core increases
at the second epoch and decreases considerably at the third (see
Fig.~\ref{Cflux}), while the component C3 becomes distinct from the
core at the same time. The flux density curve of C3 shows flare-like
behaviour, possibly corresponding to a small rise in the single-dish
flux density around July 2003. We interpret the rise in the core flux
density at the second epoch and a sudden drop at the third epoch to be
related to the ejection of C3.  Based on the flux curves in
Fig.~\ref{mhovi_flux}, it seems to be a reasonable assumption that the
flare seen in the single-dish flux density data in January 2003 is
{\it mainly} caused by a recently ejected component C2.

Observed flux density variability -- both in single-dish data and in
the VLBI components -- can be used to calculate the size of the
emission region. In the shock-in-jet model for radio variability, the
nearly symmetric flares observed at high radio frequencies
(Ter\"asranta \& Valtaoja \cite{ter94}) suggest that the flux density
variability is controlled by light travel delays across the shocked
region (Jorstad et al. \cite{jor05}, Sokolov et al. \cite{sok04})
i.e. the flux density variability time scale defined as $\tau =
dt/d\ln(S)$ corresponds to the light crossing time across the
region. This is true if the light crossing time is longer than
the radiative cooling time but shorter than the adiabatic cooling
time. Jorstad et al. (\cite{jor05}) verified this by comparing flux
density variability time scales calculated for VLBI components with
time scales of variability in apparent sizes of the same
components. They found that the majority of components had a shorter flux
variability time scale than that predicted for adiabatic expansion,
implying that at high radio frequencies the decay in flux density is
driven by radiative cooling. The above assumption allows us to
identify $\tau$ with the light-crossing time across the emission
region and hence, to estimate the size of the region, which can be
then compared with the size measured from the VLBI data to calculate
the Doppler factor.

First, we estimate the variability time scale $\tau_{\mathrm{TFD}}$
from the total flux density data. In order to make a simple
parametrisation of the flare in January 2003, we have decomposed the
flux curve into a constant baseline and separate flares consisting of
exponential rise, sharp peak and exponential decay (see Valtaoja et
al. \cite{val99} for details). Model flares are of the form
\begin{equation} \label{expfun}
\Delta S(t) = \left\{ \begin{array}{ll} \Delta S_{\mathrm{max}}
e^{(t-t_{\mathrm{peak}})/\tau_{\mathrm{TFD}}}, & t < t_{\mathrm{peak}}.
\\ \Delta S_{\mathrm{max}} e^{(t_{\mathrm{peak}}-t) / (1.3
\tau_{\mathrm{TFD}})}, & t > t_{\mathrm{peak }}.
\end{array} \right.
\end{equation}
Our best fit (dashed line) is presented in Fig.~\ref{mhovi_flux},
consisting of two small flares and a constant baseline of
11.5~Jy. Parameters of this fit are given in
Table~\ref{flare_fit}. The variability time scale for the 2003 flare
is $\tau_{\mathrm{TFD}}=175$ days. The uncertainty in
$\tau_{\mathrm{TFD}}$ is rather hard to estimate, but according to
L\"ahteenm\"aki et al. (\cite{lah99a}), a conservative upper limit for
$\sigma(\tau_{\mathrm{TFD}}) \lesssim \tau_{\mathrm{TFD}}/4$. An
angular size $a$ corresponding to $\tau_{\mathrm{TFD}}=175\pm44$~d is
given by
\begin{equation}
a = \frac{c \tau_{\mathrm{TFD}} (1+z)}{D_{\mathrm{L}}} \delta,
\end{equation}
where $D_{\mathrm{L}}$ is luminosity distance:
\begin{equation}
D_{\mathrm{L}} = \frac{c (1+z)}{H_0} \xi(\Omega_M,\Omega_{\Lambda},z),
\end{equation}
with function $\xi$ given by 
\begin{equation}
\xi(\Omega_M,\Omega_{\Lambda},z)= \int_0^z \frac{dz'}{\sqrt{(1+z')^2(1+\Omega_M z')-z
'(2+z')\Omega_{\Lambda}}},
\end{equation}
(this applies when $\Omega_M + \Omega_{\Lambda} = 1$; see Carroll et
al. \cite{car92}). For January 2003 flare, $a = (0.047\pm0.012)
\delta$~mas, which can be compared with the size of C2 measured from
the VLBI data. The full width at half maximum (FWHM) size of the major
axis of the Gaussian component C2 is $0.23\pm0.04$~mas and the axial
ratio is $\eta=0.29\pm0.06$. Taking the geometric mean of major and minor
axes and multiplying the result by 1.8, we get a size equivalent to an
optically thin sphere\footnote{The 50\%-visibility points coincide for
a Gaussian profile and for the profile of an optically thin sphere
with diameter of 1.8 times the FWHM of the Gaussian profile.},
$a_{\mathrm{VLBI}}=0.22\pm0.05$~mas. Comparison with $a$ from
variability gives $\delta = 4.7\pm1.6$ for C2.

\begin{table}
\caption{Decomposition of the single-dish flux curve into self-similar flares}
\label{flare_fit}
\centering
\begin{tabular}{cccc} \hline \hline
Flare & $t_{\mathrm{peak}}$ [year] & $\Delta S_{\mathrm{max}}$ [Jy] &
 $\tau_{\mathrm{TFD}}$ [days] \\ \hline 
1 & 2003.03 & 9.72 & $175\pm44$ \\ 
2 & 2002.06 & 6.76 & $174\pm44$ \\ \hline 
Constant baseline & & 11.5 Jy \\ \hline
\end{tabular}
\end{table}

\begin{figure}
\resizebox{\hsize}{!}{\includegraphics{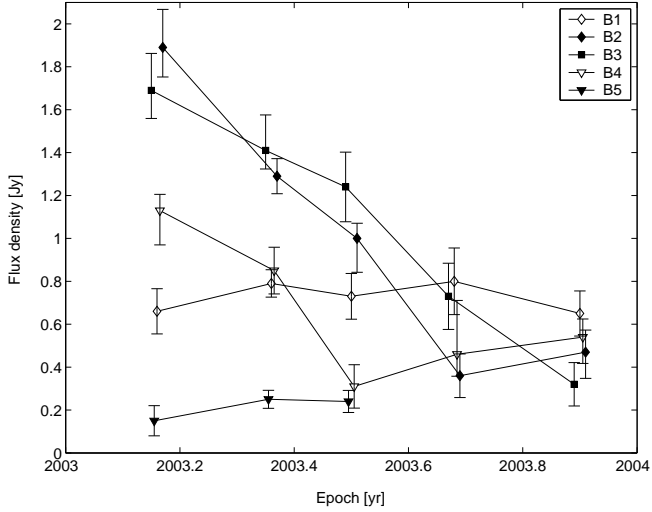}}
\caption{Flux density evolution for components B1--B5.The asymmetric
error bars are mainly due to interdependencies between flux densities
of close-by components.}
\label{Bflux}
\end{figure}

The variability time scale can also be determined solely on the basis
of flux density variability of a single VLBI component. As can be seen
from Figs.~\ref{Cflux} and \ref{Bflux}, components B2, B3, B4, C2 and
C3 show enough flux density variability that they can be used in
estimating the variability time scale. The flux density time scales,
$\tau_{\mathrm{VLBI}}$, the sizes at the time of maximum flux
densities, $a_{\mathrm{VLBI}}$ (with geometry of an optically thin
sphere assumed, i.e. the figures are multiplied by 1.8), and the
calculated Doppler factors $\delta$ are presented in
Table~\ref{d_var_table} for these components. In calculating
$\tau_{\mathrm{VLBI}}$ and $\delta$, we have taken into account the
asymmetric uncertainties in flux densities and sizes of several
components (see e.g. D'Agostini (\cite{ago04}) for a short review
about asymmetric uncertainties and their correct handling). The
figures given in Table~\ref{d_var_table} are probabilistic expected
values, not ``best values'' obtained by direct calculation, and the
uncertainties correspond to probabilistic standard deviations. 

A non-linear dependence of the output quantity $Y$ on the input
quantity $X$ in a region of few standard deviations around the
expected value of $X$ is another source of asymmetric uncertainties,
which also needs to be properly taken into account when doing
calculations with measured quantities. In this paper, when
calculating derived quantities like brightness temperature, which has
a strongly non-linear dependence on the size of the component, we have
used a second-order approximation for the error propagation formula
instead of the conventional linear one. Moreover, the non-linearity
does not only affect the errors, but it can also shift the expected
value of $Y$ with respect to the ``best value'' obtained by direct
calculation (D'Agostini \cite{ago04}). In the following, for all the
derived quantities, expected values and standard deviations are
reported instead of directly calculated values, since they are much
more useful for any statistical analysis.

\begin{table}
\begin{minipage}[t]{\columnwidth}
\caption{Variability time scales and Doppler factors from the VLBI data}
\label{d_var_table}
\centering
\renewcommand{\footnoterule}{}
\begin{tabular}{cccc} \hline \hline
Component & $\tau_{\mathrm{VLBI}}$ [days] & $a_{\mathrm{VLBI}}$ [mas] & $\delta$ \\ \hline
C2 & $110\pm30$ & $0.22\pm0.05$ & $7.5\pm2.6$ \\
C3 & $120\pm10$ & $0.12\pm0.04$ & $4.1\pm1.4$ \\
B2 & $110\pm20$ & $0.36\pm0.05$ & $12.2\pm2.8$ \\
B3 & $160\pm30$ & $0.18\pm0.04$ & $4.2\pm1.7$ \\
B4 & $100\pm30$ & $0.23\pm0.05$ & $8.6\pm3.2$ \\ \hline
\end{tabular}
\end{minipage}
\end{table}

The two variability time scales for C2, estimated from the total flux
density and from the VLBI data, differ by 60 days, which affects the
calculated Doppler factors -- from the total flux density variability
we get $\delta(C2) = 4.7\pm1.6$ as compared to $\delta(C2) =
7.5\pm2.6$ calculated from the variability in the VLBI data. A
weighted average of these, $\delta(C2)=5.5\pm1.9$, agrees well with
other Doppler factor values for 3C 273 derived elsewhere:
$\delta_{var} = 5.7$ (L\"ahteenm\"aki \& Valtaoja 1999), $\delta_{eq}
= 8.3$, $\delta_{SSC} = 4.6$ (G\"uijosa \& Daly \cite{gui96}), and
$\delta_{SSC} = 6.0$ (Ghisellini et al. \cite{ghi93}). The variability
Doppler factors of other components listed in Table~\ref{d_var_table}
vary significantly, with C3 and B3 having the smallest values,
$\delta\sim4$, and B2 having the highest,
$\delta(B2)=12.2\pm2.8$. This is not very surprising, since also the
apparent velocities of these components differ considerably.

\subsection{Brightness temperatures and equipartition}

Brightness temperature is another quantity of interest in sources like
\object{3C\,273}. From the total flux density variability, we can
estimate the variability brightness temperature (L\"ahteenm\"aki et
al. \cite{lah99a})
\begin{equation} \label{tb_var}
T_{b,\mathrm{var}} =  5.28 \times 10^{20} \cdot h^{-2} \frac{\Delta S_{\mathrm{max}
}}{\nu ^2 \tau ^2 _{\mathrm{TFD}}} (1+z) \xi(\Omega_M,\Omega_{\Lambda},z)^2,
\end{equation}
where $\nu$ is the observation frequency in GHz, $z$ is redshift,
$\Delta S_{\mathrm{max}}$ is the maximum amplitude of the outburst in
Janskys, and $\tau_{\mathrm{TFD}}$ is the observed variability
timescale in days.  The numerical factor in equation~\ref{tb_var}
comes from assuming that the component is an optically thin,
homogeneous sphere. L\"ahteenm\"aki et al. (\cite{lah99a}) estimate a
conservative upper limit to the error in $T_{b,\mathrm{var}}$ to be
$\Delta T / T < 0.50$. For the flare in January 2003,
$T_{b,\mathrm{var}} = (6.6\pm3.3) \times 10^{12}$~K.

The observed variability brightness temperature is proportional to
$\delta^3$, i.e. $T_{b,\mathrm{var}} = T_{b,\mathrm{int}}
\delta^3$. Using $\delta=5.5\pm1.9$, obtained earlier for component
C2, we get a value $T_{b,\mathrm{int}} \approx (8\pm6) \times
10^{10}$~K of intrinsic brightness temperature (the strong non-linear
dependence of $T_{b,\mathrm{int}}$ on $\delta$ shifts the expected
value of $T_{b,\mathrm{int}}$ significantly from the directly
calculated ``best value'' of $\approx 4 \times 10^{10}$~K; D'Agostini
\cite{ago04}), which is consistent with the source being near
equipartition of energy between the radiating particles and the
magnetic field. According to Readhead (\cite{rea94}) the equipartition
brightness temperature, $T_{b,\mathrm{eq}}$, depends mainly on the
redshift of the source and weakly on the observed optically thin
spectral index, the peak flux density and the observed frequency at
the peak. For the redshift of \object{3C\,273}, $T_{b,\mathrm{eq}}
\sim 8 \times 10^{10}$~K for any reasonable values of $\alpha$,
$S_{m}$ and $\nu_{m}$. This exactly matches the $T_{b,\mathrm{int}}$
calculated for the January 2003 flare, indicating that near the core
the source could have been in equipartition. Our result is consistent
with the findings of L\"ahteenm\"aki et al. (\cite{lah99a}), who have
shown that during high radio frequency flares all sources in their
sample have intrinsic brightness temperature close to the
equipartition value.

Interferometric brightness temperatures of the components are also of
interest. The maximum brightness temperature for an elliptical
component with a Gaussian surface brightness profile is
\begin{equation} \label{tb_vlbi}
T_{b,\mathrm{VLBI}}= 1.22 \times 10^{12} \cdot \frac{S_{\mathrm{VLBI}} (1+z)}{\eta a_{\mathrm{maj}}^2 \nu^2}
\end{equation}
where $S_{\mathrm{VLBI}}$ is the flux density in Janskys, $\nu$ is the
observed frequency in GHz, $z$ is redshift, $a_{\mathrm{maj}}$ is the
FWHM size of the major axis in mas, and $\eta$ is the axial ratio of
the component. In order to consider a surface brightness distribution,
which is more physical than a Gaussian, we convert the Gaussian
brightness temperature to an optically thin sphere brightness
temperature by multiplying the former by a factor of
0.67. Table~\ref{tb_vlbi_table} lists the maximum interferometric
brightness temperatures for components within 2.5~mas from the core
and with non-zero areas. Again, we have taken into account the effect
of asymmetric errors in $S_{\mathrm{VLBI}}$, $a_{\mathrm{maj}}$ and
$\eta$, as well as the non-linearity of the equation~\ref{tb_vlbi},
for error propagation.  The table also contains the average Doppler
factor and $T_{b,\mathrm{int}}$ calculated by dividing
$T_{b,\mathrm{VLBI}}$ by $\delta$. $T_{b,\mathrm{int}}$ for C2 given
in Table~\ref{tb_vlbi_table} differs from $(8\pm6) \times 10^{10}$~K
that was quoted earlier, because it is calculated using the
interferometric brightness temperature instead of variability
brightness temperature. However, the difference is not significant,
considering the overall uncertainties in measuring
$T_{b,\mathrm{int}}$

\begin{table*}
\caption{Component properties}
\label{tb_vlbi_table}
\centering
\begin{tabular}{cccccc} \hline \hline
Component & $T_{b,\mathrm{VLBI}}$ [K] & $\delta$ &
 $T_{b,\mathrm{int}}$ [K] & $\Gamma$ & $\theta$ [$\degr$] \\ \hline 
 C1 & $(7.9\pm5.5) \times 10^{10}$ & -- & -- & -- & -- \\ 
 C2 & $(1.7\pm0.7) \times 10^{11}$ & $5.5\pm1.9$ & $(3\pm2) \times 10^{10}$ 
   & $9.8\pm3.2$ & $9.8\pm2.1$ \\
 C3 & $(3.4\pm1.4) \times 10^{11}$ & $4.1\pm1.4$ & $(8\pm3) \times 10^{10}$
   & $14\pm5$ & $10.0\pm1.5$ \\
 B1 & $(4.0\pm2.3) \times 10^{10}$ & -- & -- & -- & -- \\ 
 B2 & $(2.5\pm0.7) \times 10^{10}$ & $12.2\pm2.8$ & $(2\pm1) \times 10^{9}$ 
   & $8.0\pm1.0$ & $4.3\pm1.6$ \\ 
 B3 & $(1.2\pm0.7) \times 10^{11}$ & $4.2\pm1.7$ & $(3\pm2) \times 10^{10}$ 
   & $17\pm7$ & $9.7\pm0.8$ \\ 
 B4 & $(4.0\pm1.8) \times 10^{10}$ & $8.6\pm3.2$ & $(5\pm2) \times 10^{9}$ 
   & $18\pm8$ & $6.0\pm1.2$ \\ \hline
\end{tabular}
\end{table*}

\subsection{Lorentz factors and viewing angles}

Since we have measured $\delta$ and $\beta_{\mathrm{app}}$ for a number of
components, it is possible to calculate the values of $\Gamma$ and
$\theta$. The Lorentz factor is given by equation
\begin{equation}
\Gamma = \frac{\beta_{\mathrm{app}}^2+\delta ^2+1}{2 \delta}
\end{equation}
and the angle between the jet and our line of sight by equation
\begin{equation}
\theta = \arctan \frac{2 \beta_{\mathrm{app}}}{\beta_{\mathrm{app}}^2+\delta ^2 -1}.
\end{equation}
The most reliable value of the Doppler factor was obtained for
component C2.  With $\beta_{\mathrm{app}}(C2)=8.0\pm0.7$ and
$\delta(C2) = 5.5\pm1.9$, we get $\Gamma = 9.8\pm3.2$ and $\theta =
9.8\pm2.1 \degr$.

In Table~\ref{tb_vlbi_table} we have gathered Lorentz factors and
viewing angles also for other components besides C2. The results
indicate that a transverse velocity gradient across the jet does
exist, even when the viewing angle difference is taken into
account. B2 has $\Gamma=8$, while B3 and B4 show $\Gamma\sim17$, and
even the lower limit of Lorentz factor for B4 is larger than
$\Gamma(B2)$ (see Table~\ref{gamma_min}).

%

\section{Discussion}
\object{3C\,273} is the closest quasar, and due to its proximity it
provides one of the best chances to study the physics of a strong
relativistic outflow in detail. The jet in \object{3C\,273} is
complicated, with a number of details which would be smoothed away if
the source was located at higher redshift. This complexity also makes
it hard to interpret. For instance, several interferometric studies on
the parsec scale jet of \object{3C\,273} in the 1980s and 1990s have
reported a ``wiggling'' structure with time-variable ejection angles
and component velocities (e.g. Zensus et al. \cite{zen88}, Krichbaum
et al. \cite{kri90}, B{\aa}{\aa}th et al. \cite{baa91}, Abraham et
al. \cite{abr96}, Mantovani et al. \cite{man99}).  In the following,
we will explore some possible scenarios that could lie behind the
observed curving structure, and discuss the observational evidence for
and against them.

\subsection{Precession of the jet nozzle}
\begin{figure}
\resizebox{\hsize}{!}{\includegraphics{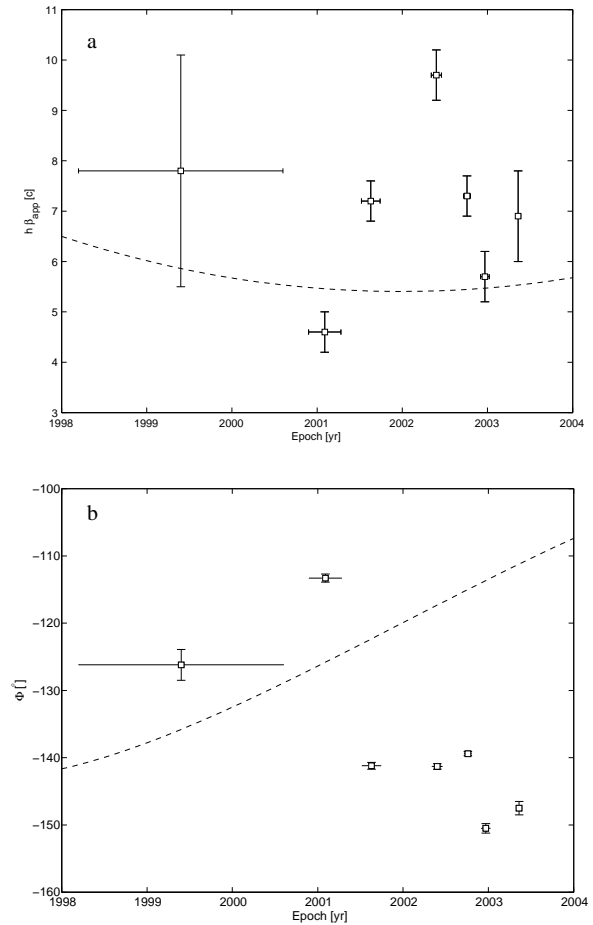}}
\caption{Apparent superluminal velocities (panel a) and direction of
motion (panel b) of the bright superluminal components ejected between
1998 and 2003 as a function of ejection epoch. The dashed curves
represent the prediction of the precession model by Abraham \& Romero
(\cite{abr99}). The horizontal error bars represent the uncertainty in
ejection epochs.}
\label{beta_vs_time}
\end{figure}

Abraham \& Romero (\cite{abr99}) proposed a simple model of a
ballistic jet with a precessing nozzle to explain the variable
ejection angles and component velocities. Their model fits well the
early VLBI data describing components ejected between 1960--1990. To
compare our more recent high frequency data with their model, we have
plotted the apparent speeds and directions of motion for the
components as a function of their ejection epochs in
Fig.~\ref{beta_vs_time} (excluding the faint B5). Overlaid in the
figure are the predicted curves from the Abraham \& Romero model. It
is clear that the component velocities and directions show much faster
variations than the precession model with a period of $\sim 15$ years
predicts. No periodicity can be claimed on the basis of our data,
although the apparent velocities may show pseudo-sinusoidal
variation. However, the directions of the components' proper motions
show no clear pattern, and the apparent direction changes as fast as
$\sim 30\degr$ in half a year.

\begin{figure}
\resizebox{\hsize}{!}{\includegraphics[angle=-90]{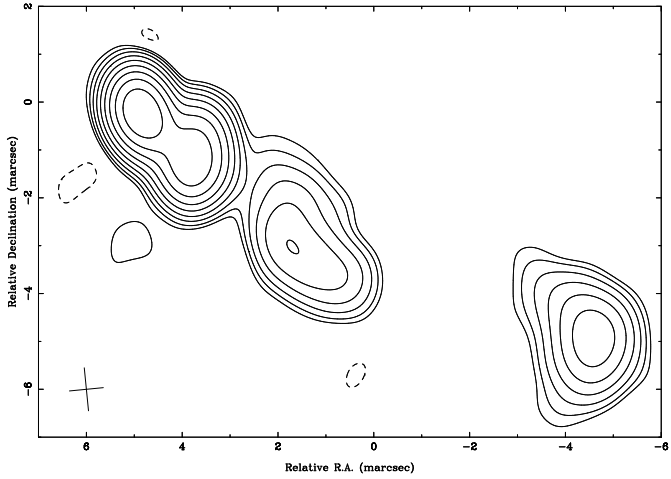}}
\caption{Image showing strongly tapered uv-data of the third epoch
  (2003.50) observation. We have tapered the 43~GHz data to a
  resolution which corresponds to observations at 10.7~GHz with a
  similar array of antennas.}
\label{taper}
\end{figure}

\begin{figure}
\resizebox{\hsize}{!}{\includegraphics{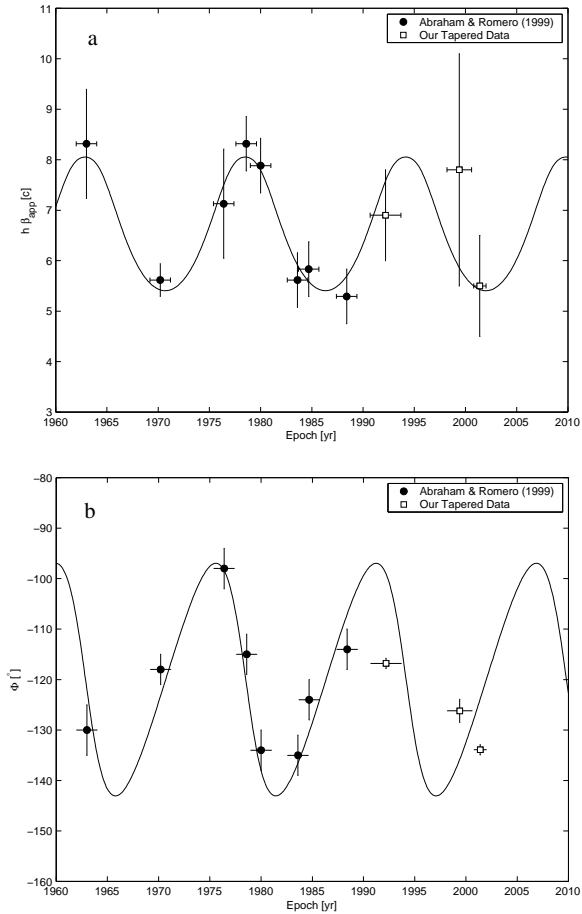}}
\caption{Apparent superluminal velocities (panel a) and direction of
motion (panel b) of the strong superluminal components as a function
of ejection epoch. We have plotted in the figure low frequency data
collected by Abraham \& Romero (\cite{abr99}) with
$\beta_{\mathrm{app}}$ scaled to the cosmology used in this paper, and
our 43~GHz data tapered to a resolution corresponding to observations
made at 10.7~GHz. The solid curves represent the precession model
of Abraham \& Romero (\cite{abr99}).}
\label{precession}
\end{figure}

There is one important difference between our data and that used by
Abraham \& Romero (\cite{abr99}) to construct their model. Most
observations cited in their paper were carried out at lower
frequencies (mainly 10.7~GHz) with significantly poorer
resolution. Many of the components in our 43~GHz maps are small and
often part of a larger emission structure, which is likely to show up
as a single component on lower resolution maps. Such a component could
show drastically different ejection angles than the individual
features in it due to averaging. We have tested whether our data could
fit the precession model of Abraham \& Romero, were it observed with
poorer resolution. The 43~GHz uv-data were tapered into a resolution
comparable to VLBA observations at 10.7~GHz and then model-fitted (an
example is shown in Fig.~\ref{taper}). In Fig.~\ref{precession} we
have plotted the apparent velocities and directions of motion for
three prominent components in the tapered data as a function of their
ejection epochs. We have also included the data from Abraham \& Romero
(\cite{abr99}), scaled to the cosmology used in this paper, and
superimposed their precession model to the figure. As can be seen, we
cannot exclude the possibility that the {\it average} jet direction
changes periodically, and that it can be approximately described by
the Abraham \& Romero model.

\subsection{Ballistic or not?}

The simplest possible model describing proper motions in a
relativistic jet is the one where knots ejected from the core move at
constant velocity following a ballistic trajectory. On the contrary,
non-ballistic motion in the jet implies a more complex physical system
with dynamics governed by e.g. magnetic fields, fluid instabilities or
interaction with the ambient medium.

Most of the components observed in our data follow, at least on a
timescale of nine months, a straight path starting from the core with
a constant velocity. However, there are a few important
exceptions. Component C1 shows a curved trajectory with slight bending
towards the north (Fig.~\ref{Cxy}). Its position at the last epoch
deviates $\sim 0.11$~mas from the straight line path extrapolated from
the first three epochs and it does not follow the path taken earlier
by components B3--B5. Follow-up observations tracking the motion of C1
will show if the bending continues and C1 becomes a ``northern
component'' with a trajectory similar to B2 or B3. The component B1 of
Jorstad et al. (\cite{jor05}) has a curved trajectory with these
characteristics, indicating that the behaviour of C1, if it actually
continues to bend, is not unique. The other two exceptions to radial
motion are components B1 and A2, which do not extrapolate back to the
core. B1 lies $\sim0.6$~mas south of B2 and moves parallel to
it. This component may have been split from B3 or B2, or it may have
changed its direction earlier -- similarly to C1.

Several studies report non-ballistic motion in \object{3C\,273}.
Kellermann et al. (\cite{kel04}) found a non-radial trajectory for one
of the components in their 2\,cm VLBA Survey monitoring, Homan et
al. (\cite{hom01}) also reported a curved path for one component, and
Jorstad et al. (\cite{jor05}) identified several knots following a
bent trajectory. Fig.~5 in Krichbaum et al. (\cite{kri00}) is a very
interesting plot, where the long term motion of 12 components is
presented in the (RA,Dec)-plane. The components show slightly curved
trajectories with both convex and concave shapes. It is apparent from
this figure that had the components been followed only for a year, one
would have described most of them as ballistic. This is in agreement
with our data, and we consider it likely that several components have
non-ballistic trajectories, although with large radii of
curvature. One more piece of evidence supporting this view comes from
the apparent opening angle of the jet. At 1--2~mas from the core, we
can estimate the apparent opening angle of the jet from the motion
directions of B2 and B3, yielding $28\degr$ in the plane of the
sky. In a 5~GHz image taken at the first epoch, the jet is
well-collimated up to 50~mas and its opening angle on the scale of
10--50~mas is significantly smaller than $28\degr$ (Paper~II;
preliminary 5~GHz data is published in Savolainen et al. \cite{sav04},
see their Fig.~1). In the kpc scale jet, the half-power width remains
constant (1.0\arcsec) from 13\arcsec to 20\arcsec along the jet
implying highly collimated outflow and, again, a small opening angle
(Conway et al. \cite{con93}). If the large observed opening angle of
$28\degr$ at 1--2~mas from the core is not due to a rare event of
changing jet direction (see Sect.~6.4), the components B2 and B3 are
not likely to continue on ballistic trajectories for long. Instead,
either collimation or increase in the viewing angle is expected to
occur, because of the smaller apparent opening angle farther
downstream. In either case, the component trajectory is not ballistic.

\subsection{Plasma instabilities}

The ``wiggling'' structure in the jet of \object{3C\,273} is present
from the subparsec (B{\aa}{\aa}th et al. \cite{baa91}) to kiloparsec
scale (Conway et al. \cite{con93}), strongly indicating that there is
a wavelike ``normal mode'' configuration in the jet. Lobanov \& Zensus
(\cite{lob01}) reported the discovery of two threadlike patterns in
the jet resembling a double helix in their space-VLBI image. They
successfully fitted the structure with a Kelvin-Helmholtz instability
model consisting of five different instability modes, which they
identify as helical and elliptical surface and body modes. The helical
surface mode in their model has significantly shorter wavelength than
its anticipated characteristic wavelength, implying that the mode is
driven externally. Lobanov \& Zensus associate this driving mechanism
with the 15-year period of Abraham \& Romero (\cite{abr99}). This is
in accordance with our observations: the average jet direction agrees
with the precession model, but individual components show much more
rapid and complex variations in speed and direction.  Lobanov \&
Zensus predicted that a substantial velocity gradient should be
present in the jet. Such a gradient was observed in this work as was
shown in the previous section.

If the magnetic energy flux in the jet is comparable to or larger than
the kinetic energy flux, the flow will, in addition to
Kelvin-Helmholtz modes, be susceptible to current-driven
instabilities. Nakamura \& Meier (\cite{nak04}) carried out
three-dimensional magnetohydrodynamic simulations that indicate that
the growth of the current-driven asymmetric (m=1) mode in a
Poynting-flux dominated jet produces a three-dimensional helical
structure resembling that of ``wiggling'' VLBI jets. Although their
simulation was non-relativistic, and thus needs confirmation of the
growth of current-driven instability modes in the relativistic regime,
the idea of current-driven instabilities in a Poynting-flux dominated
jet being behind the observed structure in \object{3C\,273} is
tempting, since the current-carrying outflow would also explain the
rotation measure gradient across the jet observed by several
groups. Asada et al. (\cite{asa02}) were the first to find a gradient
of about 80~rad~m$^{-2}$~mas$^{-1}$ at about 7~mas from the core, and
recently Zavala \& Taylor (\cite{zav05}) reported a gradient of
500~rad~m$^{-2}$~mas$^{-1}$ at approximately the same location in the
jet by using data from a higher resolution study. Attridge et
al. (\cite{att05}) performed polarisation VLBI observations of
\object{3C\,273} at 43 and 86~GHz and their data reveal a rotation
measure gradient of $\sim 10^5$~rad~m$^{-2}$~mas$^{-1}$ at 0.9~mas
from the core. These observations can be explained by a helical
magnetic field wrapping around the jet, as is expected in
Poynting-flux dominated jet models. Both Zavala \& Taylor
(\cite{zav05}) and Attridge et al. (\cite{att05}) conclude that the
high fractional polarisation observed in the jet rules out internal
Faraday rotation, and thus the synchrotron-emitting particles need to
be segregated from the region of the helical magnetic field. In a
Poynting-flux dominated jet, a strong magnetic sheath is formed
between the axial current flowing in the jet and the return current
flowing outside (Nakamura \& Meier \cite{nak04}). In this magnetic
sheath the field lines are highly twisted and it would be a natural
place to produce the observed rotation measure gradients outside the
actual synchrotron emission regions.

\subsection{Nature of region ``B''}

The most interesting structure in our images (Figs.~\ref{clean_images}
and \ref{model_images}) is the region at 1--2~mas from the core, where
there are several bright components at equal distances from the core
with vastly different directions of motion. This structure is already
present in 43 and 86~GHz images of Attridge et al. (\cite{att05})
observed at epoch 2002.35, and the authors suggest that they have
caught the jet of \object{3C\,273} in the act of changing its
direction. They also conjecture that the southern component Q7/W7 in
their data is younger and faster than the northern component Q6/W6. We
identify components Q6/W6 and Q7/W7 in Attridge et al. (\cite{att05})
with our components B2 and B3, respectively, and confirm their
conjecture about the relative age and velocity of the
components. Their first hypothesis concerning the change of jet
direction is more problematic. Supporting evidence for this change is
that the components B3--B5 and C1--C3, ejected after B2, all have
directions of motion between PA=$-140\degr$ and PA=$-150\degr$ as
compared to PA=$-113\degr$ of B2. However, the change would have had
to happen in a short time, in about six months
(Fig.~\ref{beta_vs_time}). It is hard to come up with a mechanism able
to change the direction of the entire jet so abruptly. We have
compared the positions of the components Q6/W6 and Q7/W7 of Attridge
et al. (\cite{att05}) with the positions of B2 and B3, and Q7/W7
nicely falls into the line extrapolated from the proper motion of B3,
but for Q6/W6 the situation is more complicated. Q6/W6 can be joined
with B2 by a straight line, but this path does not extrapolate back to
the core implying that Q6/W6/B2 may not follow a ballistic
trajectory. Also, the trajectory of component C1 (ejected after B2)
shows some hints of being deviated from the path taken earlier by
B3--B5, as was discussed before. If C1 continues to turn towards the
``northern track'', it may follow a trajectory similar to B1 and B2.

What possible explanations are there for the observed vastly different
directions of component motion in region ``B'' other than the changing
jet direction? One possibility is that the double helical structure of
Lobanov \& Zensus (\cite{lob01}) is already prominent at 1~mas from
the core. In their analysis there are two maxima in the brightness
distributions across the jet, and a wavelike structure would fit the
curved paths of our C1 and the component B1 of Jorstad et
al. (\cite{jor05}). Further observations following the motions of B2
and B3 are needed to confirm or disprove this explanation, but we
consider it as a simple and viable alternative to an abruptly changing
jet direction.

Another potential alternative explanation is the ``lighthouse model''
by Camenzind \& Krockenberger (\cite{cam92}), who propose that the
fast optical flux variations in the quasars may be due to a rapid
rotation of the plasma knots within the jet. The rotation changes the
viewing angle, and consequently induces a periodic variability of the
Doppler factor near the beginning of the jet. If the jet has a
non-negligible opening angle, the local rotation period increases and
the trajectories of the components asymptotically approach straight
lines as the knots move downstream. However, within a few
milliarcseconds from the core, there could still be detectable helical
motion, and the knots in the region ``B'' in \object{3C\,273} could
be components having non-negligible angular momentum and following
different paths. Again, a longer monitoring of the component
trajectories is needed to test this idea.

The jet could also broaden due to a drop in confining pressure (either
external gas pressure or magnetic hoop stress) at 1~mas from the
core. However, as the jet remains collimated in the downstream region,
we would expect the pressure drop to be followed by a recollimation,
and consequently, a jet of oscillating cross-section to form (Daly \&
Marscher \cite{dal88}). No oscillation of the jet width can be clearly
claimed on the basis of either our data or that of previous VLBI
studies. Particularly, in Jorstad et al. (\cite{jor05}), where a
broadening of the jet -- similar to the one seen in our images -- was
observed in 1998, there is no evidence of such an oscillation in the
following two years. Still, the observations are not easy to interpret
in this case, and we cannot conclusively reject a pressure drop
scenario either.

%

\section{Conclusions}

We have presented five 43~GHz total intensity images of
\object{3C\,273} covering a time period of nine months in 2003 and
belonging to a larger set of VLBA multifrequency monitoring
observations aimed to complement simultaneous high energy campaign
with the INTEGRAL satellite. We fitted the data at each epoch with a
model consisting of a number of Gaussian components in order to
analyse the kinematics of the jet. Particular attention was paid to
estimating the uncertainties in the model parameters, and the Difwrap
program was found to provide reliable estimates for positional errors
of the components.

The images in Figs.~\ref{clean_images} and \ref{model_images} show an
intriguing feature at 1--2~mas from the core, where the jet is
resolved in direction transverse to the flow. In this broad part of
the jet, there are bright knots with vastly different directions of
motion. The jet may have changed its direction here, as proposed by
Attridge et al. (\cite{att05}), although we consider it more likely
that we are seeing an upstream part of the double helical structure
observed by Lobanov \& Zensus (\cite{lob01}).

We have analysed the component kinematics in the parsec scale jet and
found velocities in the range of 4.6--13.0~$h^{-1}$~c. There is an
apparent velocity gradient across the jet with northern components
moving slower than southern ones. Thus, we can confirm the earlier
report of this gradient by Jorstad et al. (\cite{jor05}). We have also
found curved and non-radial motions in the jet, although most of the
components show radial motion on a time scale of nine months. Taking
into account other studies reporting non-ballistic motion in
\object{3C\,273} (Krichbaum et al. \cite{kri00}, Homan et
al. \cite{hom01}, Kellermann et al. \cite{kel04} and Jorstad et
al. \cite{jor05}), we consider it likely that several components in
the jet have non-ballistic trajectories, although with large radii
of curvature. This also is in accordance with the fluid dynamical
interpretation of motion in \object{3C\,273}.

By using flux density variability and light travel time arguments, we
have estimated the Doppler factors for the prominent jet components
and combined them with the apparent velocities to calculate the
Lorentz factors and viewing angles. For instance, for the newly
ejected component C2, we get an accurate and reliable value of the
Doppler factor, $\delta(C2)=5.5\pm1.9$. The Doppler factors will be
used in Paper~II together with component spectra to calculate the
anticipated amount of hard X-rays due to the synchrotron self-Compton
mechanism. The Lorentz factors obtained in this paper range from 8 to
18, and show that the observed apparent velocity gradient is not
entirely due to different viewing angles for northern and southern
components, but the southern components are indeed intrinsically
faster.

We have compared the velocities and directions of motion of the
observed components with the predictions from the precessing jet model
by Abraham \& Romero (\cite{abr99}), and we found significantly faster
variations in velocities and directions than expected from the
precession. However, after tapering our uv-data to a resolution
corresponding to 10.7~GHz data mainly used in the construction of the
Abraham \& Romero model, we obtained a moderate agreement with our
observations and the model. Thus, the {\it average} jet direction may
precess as proposed by Abraham \& Romero, but {\it individual
components} in high frequency VLBI images have faster and more random
variations in their velocity and direction.

Total flux density monitoring at 37~GHz shows that the source
underwent a mild flare in early 2003, at the time when INTEGRAL
observed it. Unfortunately, the flare was weak compared to the large
outbursts in 1983, 1988, 1991 and 1997. We can, with confidence,
identify the ejection of component C2 with the 2003 flare, and
components C1 and C3 are perhaps also connected to this event.

\begin{acknowledgements}
  This work was partly supported by the Finnish Cultural Foundation
  (TS), by the Japan Society for the Promotion of Science (KW), and by
  the Academy of Finland grants 74886 and 210338.
\end{acknowledgements}

\end{document}